\documentclass[aps,preprint]{revtex4}
\usepackage{amssymb}
\usepackage{amsmath}

\input{tcilatex}
\begin{document}

\title{Majorana states in a p-wave superconducting ring.\textit{\ }}
\author{B. Rosenstein$^{1,2}$, I. Shapiro$^{3}$, B. Ya. Shapiro$^{3}$}
\affiliation{$^{1}$Department of Electrophysics, National Chiao Tung University, Hsinchu,
Taiwan, R.O.C. }
\affiliation{$^{2}$Physics Department, Ariel University Center of Samaria, Ariel 40700,
Israel\\
$^{3}$Department of Physics, Institute of Superconductivity, Bar-Ilan
University, Ramat-Gan 52900, Israel}

\begin{abstract}
The spectrum of excitations of the chiral superconducting ring with internal
and external radii $R_{i}$, $R_{e}$ (comparable with coherence length $\xi $%
) trapping a unit flux $\ \Phi _{0}$ is calculated . We find within the
Bogoliubov-deGennes approach that there exists a pair of \textit{precisely}
zero energy states when $2k_{\perp }\left( R_{e}-R_{i}\right) /\pi $ is
integer (here $k_{\perp }$ is the momentum component in the disk plane while 
$k_{\perp }\xi >1$ ). They are not protected by topology, but are stable
under certain deformations of the system. We discuss the ways to tune the
system so that it grows into such a "Majorana disk". This condition has a
character of a resonance phenomenon.
\end{abstract}

\maketitle

\section{Introduction.}

Spin-triplet $p$-wave superfluids, both neutral, such as liquid $He^{3}$, $%
Li^{6}$ or $K^{40}$ and charged, such as superconducting material $%
Sr_{2}RuO_{4}$ and possibly heavy fermion $UPt_{3}$ have resulted in very
rich physics \cite{Leggett}. The condensate is described by a generally
tensorial complex order parameter $\Delta $ exhibiting a great variety of
the broken symmetries ground states. The broken symmetry and boundary
conditions give rise to the continuous configuration of the order parameter
as nontrivial topological excitations \cite{Volovik}. Especially interesting
is the case of the so-called topological superconductors, characterized by
the presence of electron-hole symmetry and the absence of both time-reversal
and spin-rotation symmetry. Realizations of topological $p$-wave superfluids
are chiral superconductors like $Sr_{2}RuO_{4}$, with order parameter of the 
$p_{x}\pm ip_{y}$ symmetry type \cite{Maeno} and the ABM - phase\cite%
{Leggett} of superfluid $He^{3}$ and other fermionic cold atoms, as well as
the topological superconductor $Cu_{x}Bi_{2}Se_{3}$ that produces an
equivalent pseudospin system on its surface\cite{Ong}.

Generally a magnetic field in type II superconducting films easily creates
stable line - like topological defects, Abrikosov vortices\cite{Kopnin}. In
the simplest vortex the phase of the order parameter rotates by $2\pi $
around the vortex and each vortex carries a unit of magnetic flux $\Phi _{0}$
with superfluid density depleted in the core of the size of the coherence
length $\xi $. Quasiparticles near the vortex core "feel" the phase wind by
creating a set of discrete low-energy Andreev bound state. An unpinned
vortex in an \textit{infinite} $p$-wave superconductors exhibits a
remarkable topological feature of appearance of the zero energy mode in the
vortex core \cite{Volovik99} (for each value of momentum $k_{\bot }$
perpendicular to the field direction). The zero mode represents a condensed
matter analog of the Majorana fermion \cite{Wilczek}. Its topological nature
ensures robustness against perturbations from deformations of order
parameters and nonmagnetic impurities. Due to possible applications of the
Majorana states in quantum computing it is important to ensure a relatively
large minigap $E_{mg}$ separating the Majorana states from charged
excitations. It was proposed to enlarge the minigap from \cite{Caroli} $%
\Delta ^{2}/E_{F}$ to order $\Delta $ by pinning vortices on inclusions of
small radius \cite{Shapiro11,Nori} $R_{i}\sim \xi $.

In this note we study the influence of size effects on the appearance of the
Majorana states in a ring made of the chiral superconductor. It was noted
that in finite samples of size $L>>\xi $ there generally is present there is
a \textit{pair} of "nearly" Majorana states constituting one charged fermion
degree of freedom like in 1D Kitaev model \cite{Kitaev,Alicea}. However
these states are no longer topologically protected and naively one would
expect that the degeneracy is removed with small splitting energy $\delta
E\varpropto e^{-L/\xi }$\cite{Galitsky,Radzihovsky} due to tunneling from
the core to the surface of the sample. For systems of small size $L\sim \xi $
the situation might be different. In particular it is not clear whether the
splitting always occurs at all and how it depends of the sample geometry.

To investigate the survival of Majorana states we calculate exactly the
spectrum of excitations in the chiral superconducting ring with internal and
external radii $R_{i}$, $R_{e}$ of order $\xi $ trapping a unit flux $\ \Phi
_{0}$ (see fig.1) within the Bogoliubov - deGennes approach. A surprising
finding is that although the splitting is generally of order of the bulk
energy gap $\Delta $, there exists a pair of \textit{precisely} zero energy
states for special relation $R_{e}-R_{i}\approx \pi n/2k_{\perp }$ for any
integer $n$. Then, using perturbation theory, we generalize the
consideration to other geometries.

\section{Basic equations.}

We start with the Bogoliubov - deGennes (BdG) equations for the $%
p_{x}+ip_{y} $ superconductor in the presence of a single pinned vortex. The
vector potential $\mathbf{A}$ in polar coordinates $r,\varphi $ has only an
azimuthal component $A_{\varphi }\left( r\right) $ and in the London gauge
consists of the singular part $A_{\varphi }^{s}=hc/2er$ and the regular part
of the vector potential that can be neglected for\ a type II superconductor 
\cite{Sigrist}. In the operator matrix form for a\ two component amplitude
the BdG equations read:

\begin{equation}
\left( 
\begin{array}{cc}
\hat{H}_{0} & L \\ 
L^{+} & -\hat{H}_{0}^{\ast }%
\end{array}%
\right) \left( 
\begin{array}{c}
u \\ 
v%
\end{array}%
\right) =E\left( 
\begin{array}{c}
u \\ 
v%
\end{array}%
\right) ,  \label{BdGeqs}
\end{equation}%
where for anisotropic dispersion%
\begin{eqnarray}
H_{0} &=&-\frac{\hbar ^{2}}{2m_{\bot }}\nabla _{\bot }^{2}-\frac{\hbar ^{2}}{%
2m_{\shortparallel }}\nabla _{\shortparallel }^{2}-E_{F};  \label{BdGcomp} \\
L &=&-\frac{\Delta }{k_{F}}\left\{ s\left( \mathbf{r}\right) e^{i\varphi
}\left( i\mathbf{\nabla }_{x}-\mathbf{\nabla }_{y}\right) \mathbf{+}\frac{1}{%
2}\left[ \left( i\mathbf{\nabla }_{x}-\mathbf{\nabla }_{y}\right) \left(
s\left( \mathbf{r}\right) e^{i\varphi }\right) \right] \right\} ,  \notag
\end{eqnarray}%
with $\Delta $ being the "bulk gap" of order $T_{c}$ (neglecting the small
inhomogeneity of the superfluid density within the ring). The dimensionless
profile of the order parameter $s\left( \mathbf{r}\right) $ is defined to
represent the gap function $\Delta \left( \mathbf{r}\right) =\Delta s\left( 
\mathbf{r}\right) $. In principle it should be determined self consistently,
however, for a sufficiently thin homogeneous disk we initially take $s\left( 
\mathbf{r}\right) =1$. This is justified in the bulk since the sample size
is of order of $\xi $. Although the Andreev's bound states are typically
inhomogeneous, the effect of their inhomogeneity on the order parameter is
still smaller than that of the continuum states even for small sizes \cite%
{Shapiro11}. We will later investigate the stability of the solutions with
respect to variations to $s\left( \mathbf{r}\right) $.

The equations possess the rotational and the electron-hole symmetries and
eigenstates can be found in a form:

\begin{equation}
\begin{array}{c}
u=\frac{1+i}{\sqrt{2}}f\left( r\right) e^{il\varphi }e^{ik_{\shortparallel
}z} \\ 
v=\frac{1-i}{\sqrt{2}}g\left( r\right) e^{i\left( l-2\right) \varphi
}e^{ik_{\shortparallel }z}%
\end{array}%
.  \label{ansatz}
\end{equation}%
For any angular momentum $l$ and momentum along the field $k_{\shortparallel
}$, there are radial excitation levels. In a dimensionless form Eq.(\ref%
{BdGeqs}) is written as 
\begin{eqnarray}
-\left( \frac{\partial ^{2}}{\partial r^{2}}+\frac{1}{r}\frac{\partial }{%
\partial r}-\frac{l^{2}}{r^{2}}+\frac{1}{4\gamma ^{2}}\right) f-\left( 2%
\frac{\partial }{\partial r}-\frac{2l-3}{r}\right) g &=&\varepsilon
_{lk_{\shortparallel }}f;  \label{BdGeq} \\
\left( \frac{\partial ^{2}}{\partial r^{2}}+\frac{1}{r}\frac{\partial }{%
\partial r}-\frac{\left( l-2\right) ^{2}}{r^{2}}+\frac{1}{4\gamma ^{2}}%
\right) g+\left( 2\frac{\partial }{\partial r}+\frac{2l-1}{r}\right) f
&=&\varepsilon _{lk_{\shortparallel }}g\text{,}  \notag
\end{eqnarray}%
with the dimensionless energy $\varepsilon _{lk_{\shortparallel
}}=E_{lk_{\shortparallel }}/\gamma \Delta $. Here distances are in units of $%
\xi $. In the clean limit BCS (applicable to $SrRu_{2}O_{4}$) $\xi =\hbar
k_{\bot }/m_{\bot }\Delta $, where 
\begin{equation}
k_{\bot }^{2}/2m_{\bot }=E_{F}/\hbar ^{2}-k_{\shortparallel
}^{2}/2m_{\shortparallel }\text{,}  \label{dispersion}
\end{equation}%
and for given $k_{\shortparallel }$ there is just one dimensionless
parameter 
\begin{equation}
\ \gamma =1/2k_{\perp }\xi =m_{\bot }\Delta /2\hbar ^{2}k_{\bot }^{2}<<1%
\text{.}  \label{gamma}
\end{equation}

The Ansatz, Eq.(\ref{ansatz}) was chosen in such a way that the equations
become real. In a microscopic theory of the superconductor-insulator
interface, (see \cite{DeGennes}), the order parameter rises abruptly from
zero in a dielectric, where amplitudes of normal excitations $f\left(
R_{e}\right) =g\left( R_{i}\right) =0$, to a finite value inside the
superconductor within an atomic distance $a$ from the interface, namely with
a slope $\propto 1/a$. This means that the boundary condition on the
amplitudes is consistent with a zero order parameter at the boundary point
in the self-consistency equation (see details in ref. \cite{Shapiro11} and
references therein).

\section{Majorana versus non-Majorana rings.}

Let us first determine under what conditions zero energy (Majorana) states
appear. It was shown \cite{Radzihovsky,Sigrist} that they appear only for $%
l=1$ and might contain two possible states $g_{+}=f_{+}$ or/and $g_{-}=-f_{-}
$. The corresponding equations simplify and differ just by the
transformation $r\rightarrow -r$: 
\begin{eqnarray}
\left( \frac{\partial ^{2}}{\partial r^{2}}+\left( \frac{1}{r}+2\right) 
\frac{\partial }{\partial r}+\frac{1}{r}-\frac{1}{r^{2}}+\frac{1}{4\gamma
^{2}}\right) f_{+} &=&0  \label{Majorana_eq} \\
\left( \frac{\partial ^{2}}{\partial r^{2}}+\left( \frac{1}{r}-2\right) 
\frac{\partial }{\partial r}-\frac{1}{r}-\frac{1}{r^{2}}+\frac{1}{4\gamma
^{2}}\right) f_{-} &=&0  \notag
\end{eqnarray}%
The solutions are%
\begin{equation}
f_{\pm }=e^{\mp r}\left[ c_{\pm }J_{1}\left( -r\sqrt{\frac{1}{4\gamma ^{2}}-1%
}\right) +d_{\pm }Y_{1}\left( r\sqrt{\frac{1}{4\gamma ^{2}}-1}\right) \right]
\label{Msol}
\end{equation}%
where $J_{1}$ and $Y_{1}$ are the Bessel functions. Defining the ratio $\chi
\left( r\right) =J_{1}\left( r\sqrt{\frac{1}{4\gamma ^{2}}-1}\right)
/Y_{1}\left( -r\sqrt{\frac{1}{4\gamma ^{2}}-1}\right) $ plotted in fig.2,
the boundary conditions for both $f_{+}$ and $f_{-}$ read:%
\begin{equation}
\chi \left( R_{i}\right) =\chi \left( R_{e}\right) =-d_{\pm }/c_{\pm }=\chi 
\label{ratio}
\end{equation}%
Therefore eq.(\ref{ratio}) gives a relation between $R_{i},R_{e}$ and $%
\gamma $ of the ring when two Majorana states exist \textit{simultaneously},
see the red line in fig.2 and Table I for $\gamma =1/30$ and $R_{i}=0.1$.
\bigskip 

\bigskip Table 1. External radius $R_{e}$ of the Majorana Ring with
parameters $R_{i}=0.1\xi $ and two values of $\gamma $. The radii are
identified by the integer $n$.

\begin{equation*}
\begin{array}{ccccc}
n & 1 & 2 & 3 & 4 \\ 
\gamma =1/30 & 0.256 & 0.531 & 0.742 & 0.9526 \\ 
\gamma =1/15 & 0.559 & 0.987 & 1.411 & 1.835%
\end{array}%
\end{equation*}

\bigskip 

The superfluid density%
\begin{equation}
\rho \left( r\right) =r\left\vert f_{\pm }\left( r\right) \right\vert ^{2}
\label{rho}
\end{equation}%
of the two Majorana states for $\gamma =1/30$ and $R_{i}=0.1,$ $R_{e}=0.953$
are presented in Fig. 3. One observes that due to exponentials in Eq.(\ref%
{Msol}) $f_{\pm }$ is located mostly near the internal (external) surface -
the red and green lines respectively.

Using the approximate periodicity of the Bessel functions in Eq.(\ref{Msol})
with period of $\pi $ one obtains for small $\gamma <<1$:%
\begin{equation}
R_{e}-R_{i}=2\pi n\gamma \xi =\pi n/2k_{\perp }  \label{zeroes}
\end{equation}%
where $n$ is an integer. A finite system that conforms to these conditions
will be denoted as "Majorana ring" in what follows.\ \ \ \ 

If the condition Eq.(\ref{ratio}) is violated (hence the rings will be
termed "non-Majorana") the would be Majorana $l=1$ states acquire a nonzero
energy that oscillates around zero as function of $R_{e}$ for fixed $R_{i}$
and $\gamma $. This is exemplified in Fig.4 where energies of the $l=1$
states calculated numerically are given for $\gamma =1/30$ and $R_{i}=0.1$
and a range of $\,0.76\xi <R_{e}<1.2\xi .$ The calculation utilizes the NAG
Fortran Library Routine Document F02EBF. It computes all the eigenvalues,
and optionally all the eigenvectors, of a real general matrix. One observes
that at the "Majorana geometry" the energy of the electron branch (red line)
changes sign. Similarly the hole branch has the opposite sign end the same
absolute value of energy. This exhibits a phenomenon of "level crossing" in
the BdG equation. Note that for a small sample and $R_{e}-R_{i}=\pi \left(
n+1/2\right) /2k_{\perp }$the energy of the $l=1$ states might even exceed
the gap magnitude $\Delta $, see green lines in Fig. 4.

\section{Excited states in Majorana rings.}

For a vortex in an infinite $s$-wave superconductor, when the vortices are
unpinned (freely moving) the low-lying spectrum of quasiparticle and hole
excitations is equidistant, $E_{l}=l\omega ,$ where angular momentum $l$
takes half-integer values. The "minigap" $E_{mg}$ in the $s$-wave
superfluids is of order of $\omega =\Delta ^{2}/E_{F}$ $<<\Delta $, where $%
\Delta $ is the energy gap and $E_{F}$ is the Fermi energy. In the bulk
chiral p-wave the spectrum of the low energy excitations remains
equidistant, $E_{l}=\left( l-1\right) \omega $, but now $l$ is integer\cite%
{Sigrist}.

For a Majorana ring there are excited states well below the continuum at $%
l\neq 1$ typically close to the surface. The origin of their small energy is
the vanishing of the superconducting gap on the nodes of p-wave
superconductor. Such a state with minimal energy defines the "minigap" that
protects the Majorana states from interferences due to the excitations. The
excitation energies below the threshold (namely Andreev states) were
calculated numerically for $\gamma =1/30,R_{i}=0.1$ and $R_{e}=0.98$ , see
Table 2. 

\bigskip 

\bigskip Table 2. Energy of Excited States in Majorana Ring with parameters $%
R_{i}=0.1\xi ,R_{e}=0.953\xi ,\gamma =1/30$. Energies of both the
quasiparticles ($E_{+}$) and holes ($E_{-}$) are given in units of $\Delta $.

\begin{equation*}
\begin{array}{ccccccc}
l & 1 & 2 & 3 & 4 & 5 & 6 \\ 
E_{+} & 2.6297\cdot 10^{-4} & 0.1552 & - & 0.0135 & 0.5903 & - \\ 
E_{-} & -2.6297\cdot 10^{-4} & - & -0.0437 & - & - & -0.1967%
\end{array}%
\end{equation*}

\bigskip 

Only the lowest (highest) energy $E_{+}$ ($E_{-}$) for quasiparticles
(holes) for each angular momentum is given (typically other excitations are
beyond the threshold). At certain $\ l$ there are no Andreev states.
\bigskip In the case of the small ring considered here one observes that the
minigap $E_{mg}=0.0135$ appears at $l=4$.

The energy is just a fraction of $\Delta $, still an order of magnitude
larger than $\Delta ^{2}/E_{F}$ for\ $Sr_{2}RuO_{4}$. The superfluid density 
$\rho $ defined in Eq.(\ref{rho}) corresponding to its wave function is
presented in Fig.3 (blue line). The wave function even for such a small ring
is peaked near the outer surface. To understand a relatively small value of
the minigap compared to the one that protects the core Majorana states in
the infinite system considered \cite{arXiv}, $E_{mg}=0.25\Delta \,$, we have
simulated finite large rings with $R_{e}=10,20$. Of course the order
parameter is no longer constant over such sizes and we have used $\Delta
\left( r\right) =\Delta \tanh \left( r/\xi \right) $ for the order parameter
distribution. One observes that there are surface modes with energies $%
E\left( 20\xi \right) =0.01\Delta $, $E\left( 10\xi \right) =0.01\Delta $.
Therefore the energy of these modes is almost independent of $R_{e}$.

\section{Stability with respect to change of the order parameter
distribution.}

The quantized geometry of the Majorana ring is not expected to be robust
against perturbations like disorder, shape change, magnetic field
distribution. We conjecture that varying parameters of the system, like the
order parameter that depends on local temperature etc., one can still tune
the system into a Majorana ring. To support this conjecture we calculate in
perturbation theory the correction to the Majorana solution of the simple
model Eq.(\ref{Msol}) in which the order parameter was approximated by a
constant. Now we assume that the order parameter, in addition to a
homogeneous rotation of the phase, depends on location $s\left( \mathbf{r}%
\right) =s\left( r,\varphi \right) =1+\psi \left( r,\varphi \right) $.

The only component of the BdG operator in Eq.(\ref{BdGeqs}) that is
corrected is the off diagonal:$L=-\frac{\Delta }{k_{F}}\left\{ s\left( 
\mathbf{r}\right) e^{i\varphi }\left( i\mathbf{\nabla }_{x}-\mathbf{\nabla }%
_{y}\right) \mathbf{+}\frac{1}{2}\left[ \left( i\mathbf{\nabla }_{x}-\mathbf{%
\nabla }_{y}\right) \left( s\left( \mathbf{r}\right) e^{i\varphi }\right) %
\right] \right\} $

\begin{eqnarray}
L\left[ s\left( r,\varphi \right) \right] &=&-\frac{\Delta }{2k_{F}}\left\{
2se^{i\varphi }\left( i\mathbf{\nabla }_{x}-\mathbf{\nabla }_{y}\right) 
\mathbf{+}\left[ \left( i\mathbf{\nabla }_{x}-\mathbf{\nabla }_{y}\right)
se^{i\varphi }\right] \right\}  \label{Ls} \\
&=&-\frac{i\Delta }{2k_{F}}e^{2i\varphi }\left\{ 2s\left( \frac{\partial }{%
\partial r}+\frac{i}{r}\frac{\partial }{\partial \varphi }\right) \mathbf{+}%
\frac{1}{r}\left( i\frac{\partial s}{\partial \varphi }-s\right) \mathbf{+}%
\frac{\partial s}{\partial r}\right\} .
\end{eqnarray}%
The energy correction to the $l=1$ Majorana states given in Eqs.(\ref{Msol})
to leading order in $\psi $ is

\begin{equation}
E_{M}^{ab}\bigskip =\left\langle f_{a}\left\vert H\left[ s\right] -H\left[ 1%
\right] \right\vert f_{b}\right\rangle =\int 
\begin{array}{cc}
u_{a}^{\ast } & v_{a}^{\ast }%
\end{array}%
\left( 
\begin{array}{cc}
0 & L\left[ s\right] -L\left[ 1\right] \\ 
L\left[ s\right] ^{+}-L\left[ 1\right] ^{+} & 0%
\end{array}%
\right) 
\begin{array}{c}
u_{b} \\ 
v_{b}%
\end{array}
\label{Eab}
\end{equation}%
where $a,b=\pm $ (for the internal or external surface Majorana states of
the unperturbed system). Using

\begin{equation}
L\left[ s\right] -L\left[ 1\right] =-\frac{i\Delta }{k_{F}}e^{2i\varphi
}\left\{ \psi \left( \frac{\partial }{\partial r}+\frac{i}{r}\frac{\partial 
}{\partial \varphi }\right) \mathbf{+}\frac{1}{2r}\left[ i\frac{\partial
\psi }{\partial \varphi }-\psi \right] \mathbf{+}\frac{1}{2}\frac{\partial
\psi }{\partial r}\right\} \text{,}
\end{equation}%
the diagonal elements are%
\begin{equation}
E_{M}^{++}\bigskip =E_{M}^{--}=i\int_{\varphi ,r}rf_{+}\left( r\right)
e^{-i\varphi }\left( L\left[ s\right] -L\left[ 1\right] \right) \left[
f_{+}\left( r\right) e^{i\varphi }\right] +cc\bigskip =0\text{.}
\label{diag}
\end{equation}%
The off diagonal read%
\begin{equation}
E_{M}^{-+}\bigskip =E_{M}^{+-}=-\frac{\Delta }{\pi k_{F}}\int_{\varphi
r}\psi \left( r,\varphi \right) rf_{-}f_{+}\left( r\right)  \label{def}
\end{equation}

The condition for stability of the Majoranas therefore is: $\int_{\varphi
r}s\left( r,\varphi \right) rf_{-}f_{+}=\int_{\varphi r}rf_{-}f_{+}$. For
any distribution that obeys this restriction, like for example $s\left(
r,\varphi \right) =1+s_{2}\left( \varphi \right) $ with $\int_{\varphi
}s_{2}\left( \varphi \right) =0$, there is no splitting of the Majorana
states. In particular this indicates that small deviations from the
"Majorana ring" condition can be tuned away by controlling for example the
value of the order parameter.

\section{Discussion and conclusions. \protect\bigskip}

To summarize, the spectrum of excitations of a single vortex in a chiral
superconducting ring with internal and external radii $R_{i}$, $R_{e}$
comparable with the coherence length $\xi $ was calculated. There is a pair
of \textit{precisely} zero energy states for $R_{e}-R_{i}=\pi n/2k_{\perp }$
for any integer $n$ for $k_{\perp }\xi >1$ where $k_{\perp }=\sqrt{2m_{\bot
}E_{F}/\hbar ^{2}-m_{\bot }/m_{\shortparallel }k_{\shortparallel }^{2}}$ is
the momentum in direction perpendicular to the magnetic field. Therefore a
certain combination of geometrical factors, order parameter (dependent on
temperature and material parameters) and magnetic field distribution make a
"Majorana ring". They are not protected by topology, but are stable under
certain deformations of the system. The condition has a character of a
resonance phenomenon.

The pair of exact Majorana states constitutes a qubit and might offer a
promising method of the fault-tolerant quantum computation \cite{Kitaev}. It
is separated by a minigap of order $\Delta /70$ from charged Andreev surface
states, still order of magnitude larger than the Caroli,Matricon, deGennes
minigap \cite{Caroli} $\Delta ^{2}/E_{F}$ for the superconductor $%
Sr_{2}RuO_{4}$ with $\Delta =2K,\ E_{F}=10^{3}K$. These states are generally
ignored in other qubit proposals based on pair of Majorana states that
belong to two different vortices. This configuration was comprehensively
studied by Matsishita and Machida \cite{Machida} under assumption that the
distance between the vortices $L$ is much larger than $\xi $, so that
tunneling between them can be treated as a perturbation. They found, using
variational state made of the core Majorana states of each of the vortices,
that the splitting energy of the two nearly Majorana states oscillates $%
E_{\pm }\sim \Delta \sqrt{k_{F}L}\cos \left( k_{F}L+\pi /4\right) e^{-L/\xi
} $. Therefore at $L=\pi \left( n+1/4\right) /k_{F}$ one would get a
degeneracy. This is reminiscent of our formula with two important
differences. First our splittings are generally of order $\Delta $, not $%
\Delta \sqrt{k_{F}L}e^{-L/\xi }$ and second, more importantly, the Majorana
states highly overlap, so that $n$ is small, while for two vortices $%
n>Lk_{F}>>1$.

Practical proposals will depend on the thickness of the film. In the 2D
limit $k_{\perp }=k_{F}$ (the condition is determined below) and the
Majorana condition is $R_{e}-R_{i}=\pi n/2k_{F}$. For $Sr_{2}RuO_{4}$ with 
\cite{Maeno}, $\xi =65nm$ the ring width should be a quantized in $\pi
/2k_{F}=4nm\approx \xi /30$ with minigap of order of $1K$. Moreover
deviations from the Majorana condition lead to a very sharp splitting of
order $\Delta $, see Fig.3. When the thickness of the film is $D$ new
channels open for the Andreev bound states: $k_{\shortparallel }=2\pi
n_{\shortparallel }/D$, where $n_{\shortparallel }$ is integer. The new
condition on the width becomes: $R_{e}-R_{i}=\frac{\pi }{2k^{\ast }}\left[
1-\alpha \left( 2\pi n_{\shortparallel }/Dk^{\ast }\right) ^{2}\right]
^{-1/2}$, where $\alpha =m_{\bot }/m_{\shortparallel }=0.03$ is anisotropy
and $k^{\ast }=\sqrt{2m_{\bot }E_{F}}/\hbar $. The second channel $%
n_{\shortparallel }=1$ will enter for $R_{e}-R_{i}\sim \xi $ at width $%
D=2\pi \alpha ^{1/2}/k^{\ast }=5nm$. Therefore the film is 2D only when it
is thinner than that, while for thicker films, when several new channels are
open, one of them can harbor Majorana states.

\textit{Acknowledgements.} We greatly appreciate discussions with L.N.
Bulaevskii, M. Meidan, B.Y. Zhu, and M. Lewkowicz. B.Ya.S. and I.S.
acknowledge support from the Israel Scientific Foundation.

\newpage

\bigskip

Figure captions

\bigskip

Fig.1

A p-wave superconducting mesoscopic disk with internal and external radii $%
R_{i}$ and $R_{e}$ subject to a magnetic field perpendicular to the disk.

Fig.2

The resonance condition for the Majorana disk. The function $\chi \left(
r\right) $ for two values of the parameter $\gamma =1/2\xi k_{\perp }$, $%
\gamma =1/30$ (blue), and $\gamma =1/15$ (red). To obey the resonance
condition the internal and external radii $R_{i}$ and $R_{e}$ should obey $%
\chi \left( R_{i}\right) =\chi \left( R_{e}\right) $.

Fig.3

Superfluid density (defined in Eq.(\ref{rho})) of low energy states in
Majorana ring with internal and external radii $R_{i}=0.1\xi $ and $%
R_{e}=0.953\xi $ respectively and $\gamma =1/30$. The two Majorana $l=1$
states are the internal or core state (red) and external surface state
(green) and the lowest energy excited charged state with $l=4$ (blue). The
excited state clearly resembles the surface states of p-wave superconductors.

Fig.4.

Violation of the resonance condition. The Majorana states are split into
positive and negative charged branches. The splitting energy in mesoscopic
samples (like the one presented with $R_{i}=0.1\xi $ and $R_{e}$ in the
range $0.7\xi -1.3\xi $) can be as large as the superconducting gap (green
line) at $R_{e}=0.8$. One observes that for geometries far from the Majorana
ring condition, $R_{e}=0.742,$ $0.953$ (for $\gamma =1/30$). Note the
oscillation of splitting energy of quasiparticle and hole.


\bigskip

\newpage 

\begin{thebibliography}{99}
\bibitem{Leggett} Legget A.J., \textit{Quantum Liquid: Bose Condensation and
Cooper pairing in Condensed-Matter Systems}, (Oxford University Press,
Oxford ) 2006.

\bibitem{Volovik} Salomaa M. M. and Volovik G. E. , \textit{Rev. Mod. Phys}. 
\textbf{59}, (1987) 533 ; Volovik G. E., \textit{Universe in a Helium Droplet%
} (Oxford University Press, London) 2003.

\bibitem{Maeno} Maeno Y. \textit{et al.} \textit{\ Nature} \textbf{372}
(1994) 532 ; Mackenzie A. P. and Maeno Y., \textit{Rev. Mod. Phys.} \textbf{%
75} (2003) 657.

\bibitem{Ong} Hor Y. S. \textit{et al}, \textit{Phys. Rev. Lett.} \textbf{104%
} (2010) 057001 ; Wray L. A., \textit{Nat. Phys.} \textbf{6} (2010) 855.

\bibitem{Kopnin} Kopnin N., Vortices in type-II superconductors: Structure
and Dynamics", (Oxford University Press, Oxford) 2001; Rosenstein B. and Li
D.P. \textit{Rev. Mod. Phys.} \textbf{82} (2010) 109.

\bibitem{Volovik99} G. E. Volovik, \textit{JETP Lett.} \textbf{70} (1999)
609.

\bibitem{Wilczek} F. Wilczek, \textit{Nat. Phys.} \textbf{5} (2009) 614.

\bibitem{Caroli} Caroli C., de Gennes P.G. and Matricon \textit{J.} \textit{%
Phys. Lett.,} \textbf{9} (1964) 307.

\bibitem{Shapiro11} Mel'nikov A. S. , Samokhvalov A. V. , and Zubarev M. N. 
\textit{Phys. Rev.B }\textbf{79} (2009) 134529; Rosenstein B. \textit{et al} 
\textit{Phys. Rev. B }\textbf{84}, 134521 (2011).

\bibitem{Nori} Rakhmanov A. L., Rozhkov A. V. and Nori F. \textit{Phys. Rev.B%
} \textbf{84} (2011) 075141.

\bibitem{Kitaev} Kitaev A. \textit{Ann. Phys.} \textbf{303} (2003) 2 ;
Ivanov D. A. \textit{Phys. Rev. Lett. 86} (2001) 268 ; Stone M. and Chung
S.-B. \textit{Phys. Rev. B }\textbf{73}, (2006) 014505; Tewari S. \textit{et
al., Phys. Rev. Lett. }\textbf{98}, (2007) 010506 ; C. Nayak \textit{et al} 
\textit{Rev. Mod. Phys.} \textbf{80} (2008) 1083.

\bibitem{Alicea} Alicea J. \textit{Phys. Rev.B }\textbf{81 }(2010)\textbf{\ }%
185318.

\bibitem{Galitsky} Cheng M, \textit{et al Phys. Rev. Lett.} \textbf{103}
(2009) 107001.

\bibitem{Radzihovsky} Gurarie V and Radzihovsky L, \textit{Phys. Rev. B }%
\textbf{75} (2007) 212509.

\bibitem{Sigrist} Matsumoto M and Sigrist M, \textit{J. Phys. Soc. Jap.} 
\textbf{68} (1999) 724; Fujimoto S. \textit{Phys. Rev. B }\textbf{77},
(2008) 220501(R); Sato M. and Fujimoto S, \textit{Phys. Rev. B }\textbf{79}
(2009) 094504.

\bibitem{DeGennes} deGennes P. G. Superconductivity of metalls and alloys,
(W.A. Benjamin inc., New York) 1966.

\bibitem{Machida} Mizushima T. and Machida K. \textit{Phys. Rev.A }\textbf{82%
} (2010) 023624.

\bibitem{arXiv} Rosenstein B. \textit{et al} arXiv:1210.1835\ (2012).
\end{thebibliography}
\end{document}